\pdfoutput=1

\documentclass[11pt]{article}

\usepackage[]{template/acl}

\usepackage{times}
\usepackage{latexsym}

\usepackage[T1]{fontenc}

\usepackage[utf8]{inputenc}

\usepackage{microtype}

\usepackage{inconsolata}

\usepackage{soul}
\usepackage[ruled]{algorithm2e}
\SetKwComment{Comment}{/* }{ */} 
\usepackage{graphicx}
\graphicspath{{./figure/}}
\usepackage{bm}
\usepackage{amsmath,amsfonts,amssymb}
\usepackage{multirow}
\usepackage{placeins}
\usepackage{subcaption}
\usepackage{comment}
\usepackage{array}
\usepackage{stfloats}
\usepackage{booktabs}
\usepackage{mdframed}
\usepackage{xcolor}

\definecolor{prompt_background}{HTML}{f4faed}

\title{Improving Tool Retrieval by Leveraging Large Language Models for Query Generation}

\author{
Mohammad Kachuee, Sarthak Ahuja, Vaibhav Kumar, Puyang Xu, Xiaohu Liu\\
Amazon Alexa AI\\
\{kachum, sarahuja, kvabh, puyax, derecliu\}@amazon.com
}
\date{}

\begin{document}
\maketitle

\begin{abstract}
Using tools by Large Language Models (LLMs) is a promising avenue to extend their reach beyond language or conversational settings. The number of tools can scale to thousands as they enable accessing sensory information, fetching updated factual knowledge, or taking actions in the real world. 
In such settings, in-context learning by providing a short list of relevant tools in the prompt is a viable approach. To retrieve relevant tools, various approaches have been suggested, ranging from simple frequency-based matching to dense embedding-based semantic retrieval. However, such approaches lack the contextual and common-sense understanding required to retrieve the right tools for complex user requests.
Rather than increasing the complexity of the retrieval component itself, we propose leveraging LLM understanding to generate a retrieval query. Then, the generated query is embedded and used to find the most relevant tools via a nearest-neighbor search.
We investigate three approaches for query generation: zero-shot prompting, supervised fine-tuning on tool descriptions, and alignment learning by iteratively optimizing a reward metric measuring retrieval performance.
By conducting extensive experiments on a dataset covering complex and multi-tool scenarios, we show that leveraging LLMs for query generation improves the retrieval for in-domain (seen tools) and out-of-domain (unseen tools) settings.
\end{abstract}

\section{Introduction}
Large Language Models (LLMs) have shown great promise in common sense language understanding, conversational fluency, and reasoning~\cite{bubeck2023sparks}. Recently, various studies explored extending such capability beyond language or conversational medium to leveraging it for using tools that are often accessible via Application Programming Interfaces (APIs)~\cite{patil2023gorilla,qin2023tool,qin2023toolllm,li2023api}. 

To introduce the tool use capability when dealing with a large number of APIs, in-context learning (ICL) provides a scalable method by presenting a set of available tools within the prompt context, and using the LLM for making the final API selection and argument filling~\cite{hudevcek2023large}. In such settings, due to prompt length and compute limitations, retrieving a short list of relevant APIs (typically less than 10) from the pool of thousands of APIs to present in the context is a key step in the pipeline. The set of retrieved APIs needs to be high-recall, i.e. it should include all APIs required for accomplishing the desired goal.

Various retrieval methods have been used for such task, including bag-of-words and frequency-based methods such as BM25 and TF-IDF that are easy to implement and computationally efficient but 
lack semantic understanding. Alternatively, embedding-based dense retrievers are generally based on sentence embeddings (e.g., SBERT~\cite{reimers2019sentence}) and nearest neighbor search (e.g., cosine similarity)~\cite{izacard2021unsupervised,johnson2019billion,yates2021pretrained}. In the typical dense retrieval setting, an index is built on API descriptions provided by developers as keys, and the user's utterance is used as the query. The key and queries can be embedded with a common encoder or separate encoders (aka dual encoders)~\cite{zhao2022dense}.

While embedding-based retrieval methods are more robust to language variations than frequency-based methods, they still lack contextual and common-sense understanding compared to the state-of-the-art LLMs. Moreover, simply relying on nearest neighbor matching is susceptible to getting mislead by extra information present in the utterance, especially for cases that require understanding the user's intention, tools, and ambiguities present in real-world interactions.

In this study, we propose leveraging LLMs to dynamically generate tool retrieval queries based on the user's utterance, where each query describes a tool required to accomplish the request. Then, such queries are used for dense retrieval. Our approach relies on the common-sense and contextual understanding of LLMs 
rather than increasing the complexities of the retrieval components.

The idea of using LLMs to improve retrieval has been studied in the literature before. For example, using LLMs to generate augmentation data for enriching the retrieval index~\cite{chowdhury2022novelty}. Alternatively, to improve the embedding models feedback from LLMs attention to the retrieved items is used to generate supervision signal to train stronger embeddings for the downstream task~\cite{rubin2021learning,li2023unified}. While these methods offer advantages over vanilla dense retrieval, the outcome is a more complex retrieval layer that still cannot match the commonsense understanding of LLMs.
Instead, in this paper, we focus on leveraging the LLM's capability 
and explore zero-shot prompting, supervised fine-tuning, and alignment learning 
approaches. Based on the experimental results, LLM-generated queries substantially improve tool retrieval in settings where a dataset of tools is available at the training time (in-domain) and when interacting with unseen tools (out-of-domain).

\section{Problem Settings}
\subsection{API Retrieval}
A basic embedding-based dense retriever consists of two main components: (a) an embedding model to map natural language to fixed-length vector representations, and (b) an index retrieval mechanism to get the most similar items given a new sample. For the case of API retrieval, typically, developers provide the description of their API in natural language which can be used to generate index keys. A user's utterance can be directly considered as a semantic retrieval query. 

\begin{figure*}[t]
    \centering
        \includegraphics[width=0.8\linewidth]{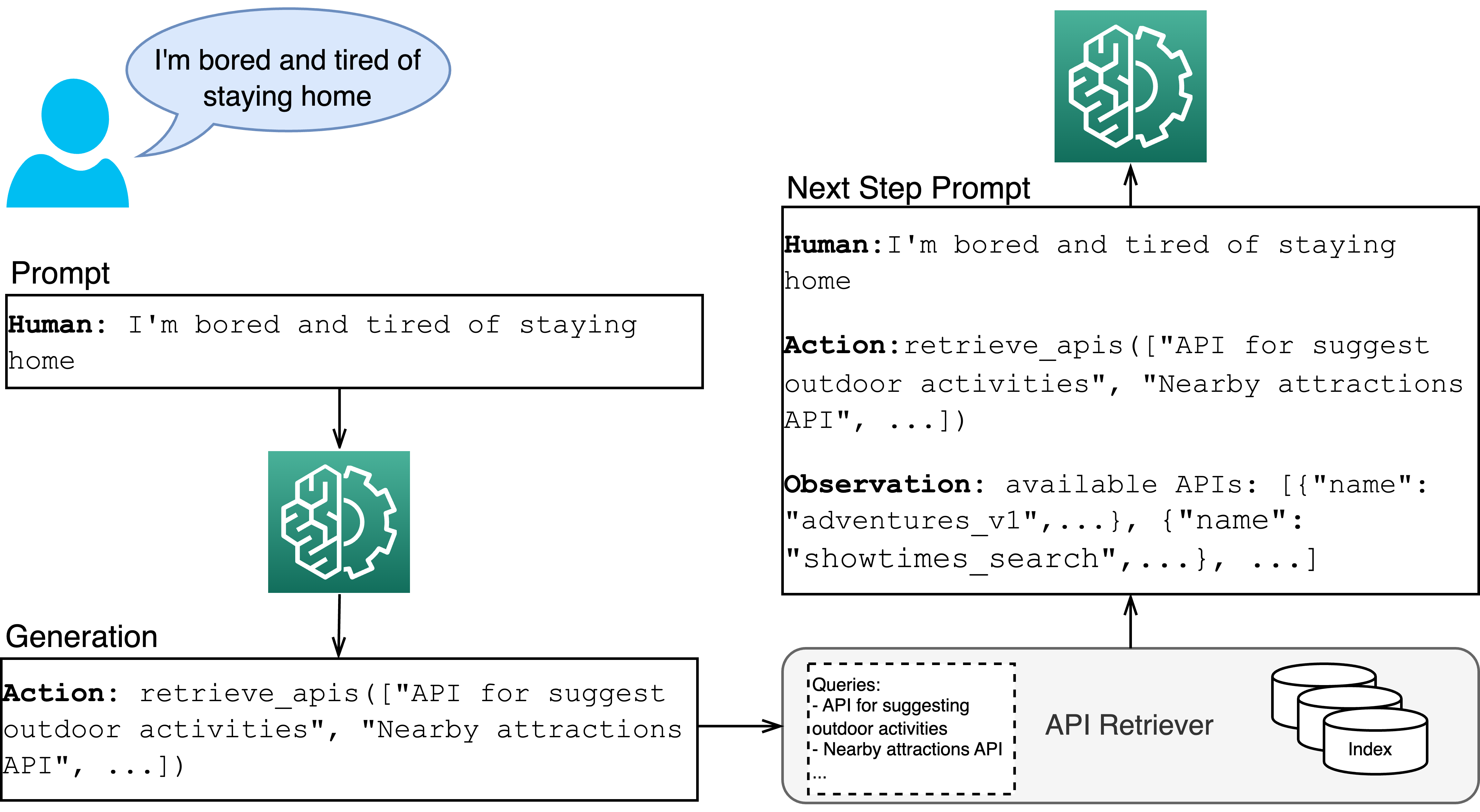}
        \caption{An illustration of leveraging LLMs commonsense and contextual understanding to generate queries for tool retrieval. The steps before and after retrieval are similar to a typical in-context learning setup not shown here.}
        \label{fig:example_flow}
\end{figure*}

Alternatively, to handle complex/contextual cases, LLM's capability to understand the conversational context can be leveraged to decompose requests and generate queries that are most suited for retrieval. Figure~\ref{fig:example_flow} shows an example flow for the query-based API retrieval. Here, the LLM reasons over the request and creates queries to be used for retrieval. Ultimately, the retrieved APIs are presented to the LLM to plan the next actions.

In this paper, we consider the problem of retrieving APIs for complex requests. A complex request requires a higher level of common-sense and semantic understanding than what is achievable via simple dense retrieval. Complex requests are often ambiguous or involve invoking multiple APIs. For example, take \textit{``I'm bored and tired of staying home. Literally watched tv all day. Give me some ideas what to do''}. In this example, a potential solution is to retrieve a list of APIs that are related to outdoor activities; however, simple dense retrieval may retrieve APIs for watching TV shows!

More formally, for a given user utterance, the tool retriever's task is to propose a ranked list of APIs, where the size of the list is denoted by $|\bm{h}|=k$. Also, when available, we are provided with a ground-truth set of relevant APIs $|\bm{y}|=n$ where $n$ $(1 \leq n \leq k)$ is the total number of relevant items for the specific sample. 

\subsection{Retrieval Metrics}
To evaluate the relevance of the retrieved results, we define three primary metrics: Recall at rank $X$ ($Recall@X$), Multiple Mean Reciprocal Rank ($MMRR$), and Mean Average Precision ($MAP$).

Assuming $\Gamma(\bm{h}_i,\bm{y})$ is an indicator function that is set to one if $\bm{h}_i$ is in the set of relevant items ($\bm{y}$) and zero otherwise, we define $Recall@X$ as:

\begin{equation}
    Recall@X = \frac{1}{n} \times \sum_{i=1}^{X} \Gamma(\bm{h_i},\bm{y}) \; .
\end{equation}

Here, $Recall@X$ is reporting for a cut-off at $X$, what percentage of relevant items would be retrieved in the set of retrieved items.

We introduce Multiple Mean Reciprocal Rank ($MMRR$) as a generalization of the Mean Reciprocal Rank \cite{radev2002evaluating} to consider cases with multiple relevant items are present:

\begin{multline}
    MMRR = \\
\resizebox{\linewidth}{!}{$
\frac{\frac{n}{2}}{\frac{1}{n} [\sum_{i=1}^{k} i \Gamma(\bm{h_i}, \bm{y}) + (k+1)(n-\sum_{i=1}^{k} \Gamma(\bm{h_i}, \bm{y}))]} .
    $}
\label{eq:mmrr}
\end{multline}

The numerator of~\eqref{eq:mmrr} is the average rank position for perfect retrieval of $n$ items. In the denominator, we compute average rank position for retrieved relevant items while clipping the tail by considering any missing item in the set of $k$ retrieved items to appear at rank $k+1$. Intuitively, $MMRR$ measures the average rank where the relevant items appear in the ranked list normalized by the best case where all top results are relevant items. With this definition, $MRRR$ reaches to one for perfect retrieval of all relevant items and is gradually reduced when the retrieval quality degrades. 

Mean Average Precision ($MAP$) is defined based on computing a finite sum of precision for the ranked list at each position~\cite{zhu2004recall}:

\begin{equation}
    MAP = \frac{1}{n} \times  \sum_{i=1}^{k} \frac{\sum_{j=1}^{i} \Gamma(\bm{h_j}, \bm{y})}{i} \times \Gamma(\bm{h_i}, \bm{y}) ,
\end{equation}
where the first term in the outer summation is precision at rank $i$. $MAP$ for perfect ranking takes the value of one, gets smaller values as relevant items appear further in the retrieved list, and reaches zero when no relevant item is retrieved.

\section{LLM-Based Query Generation}
\label{sec:LLM-Based Query Generation}
In this section, we explore three approaches to leverage LLMs for retrieval query generation including zero-shot prompting, supervised fine-tuning for API description generation, and alignment learning for optimizing the end-to-end retrieval performance. We provide additional details about the implementation, hyper-parameters selection, and ablation studies in the appendices.

\subsection{Zero-Shot Prompting}
As a simple baseline, we prompt the 13B parameter LLaMA~\cite{touvron2023llama} model to generate a description of tools required to address the user's request. We consider this method as zero-shot since there is no use of API information or task supervision therefore it can be directly applied to unseen APIs. Figure~\ref{fig:desc_prompt} shows the prompt template used for this method.

\begin{figure}[t]
\begin{mdframed}[backgroundcolor=prompt_background,linewidth=0.0pt]
\texttt{\small
Given a request by user (Human), generate the description of an API(s) that can be used to address the request.\\
Try to decompose the request to a set of descriptions for API(s) that can help handle the request.\\
Do NOT respond to the Human and just describe the API(s) that can help.\\
Use new line to separate multiple descriptions. Each description should be less than 20 words.\\
Return at most 5 descriptions (lines).\\
Do not provide any additional explanation or examples, return just a set of API descriptions.\\
\\
Human: <user request>
\\
Answer:
}
\end{mdframed}
\caption{Prompt format used for the tool description generation experiments.}
\label{fig:desc_prompt}
\end{figure}

In our early experiments with simpler prompts, we observed that the pre-trained model is inclined to attempt answer the Human directly rather than following the query generation task. We were able to mitigate this type of hallucination to some extent by emphasizing \textit{``Do NOT respond to the Human and just describe the API(s) that can help''} in the final prompt shared above.

Additionally, we found that in many cases the generated response is formatted differently than what is expected. For example, the output is formatted as a numbered list, or additional information is provided before (e.g. “Sure, I can...”) and after (e.g. “These APIs...”) the list output. To address these, we devised a set of heuristics in the output parser logic to skip invalid starting characters in the list and explanatory phrases outside the list to ensure that the right outputs are captured.

Since intent classification has been traditionally used in dialogue systems for skill selection~\cite{kachuee2022scalable}, we also conducted additional experiments instructing the LLM to generate a list of user intents rather than describing the required APIs. Note that intents provide a different abstraction of user requests than tools. In general, an intent can be potentially served by multiple tools or a tool can handle multiple intents. While the intent generation method shows marginal regressions over the tool description generation method, we found it it be less inclined to hallucination. See Appendix~\ref{sec:appendix_itent} for more details.

\subsection{Supervised fine-tuning}
To address the challenges of zero-shot prompting, and assuming we have a dataset of user utterances paired with relevant API documents, we can finetune the model for the query generation task. Specifically, we reused the prompt template from zero-shot experiments and considered the list of ground-truth relevant API descriptions as the generation target label.

Based on our initial experiments, we found that keeping the instruction prompt, limiting training to one epoch with weight decay regularization, and only computing the loss for generated tokens improved convergence and reduced overfitting.

\subsection{Alignment Learning}
While supervised fine-tuning alleviates the issues with hallucination and output inconsistency, teacher forcing (i.e., training objective enforcing generated sequence to match the target sequence) to regenerate descriptions for a specific training dataset may result in overfitting on the seen set of examples and APIs. This causes unreliable behaviors for APIs that are not seen during the training process. Note that training on a specific set of APIs may teach the LLM to try to match the current set of APIs for any new request, regardless of the availability of additional tools at the time of inference.

Apart from this, the API descriptions are typically provided by individual developers, often do not follow any strict format/content protocol, and may contain extra/irrelevant information. This can potentially bias the finetuned model and mislead the retrieval process. In other words, even perfectly generating a list of API descriptions does not necessarily result in a desirable behavior in terms of relevant API retrieval, especially when targeting out-of-domain applications.

To address these issues, we devise an alignment training scheme based on rejection sampling~\cite{bai2022constitutional} to teach LLM to generate queries that result in the best retrieval performance. Rather than directly forcing the model to generate a particular target sequence, we define a reward metric measured based on the downstream retrieval performance, and then encourage high-reward generations in an iterative alignment learning loop.

\begin{algorithm}[t]
\small
\caption{Alignment Learning Process}
\label{alg:alignment}

\DontPrintSemicolon
\SetKwInOut{Input}{input}
\SetKwInOut{Output}{output}

\Input{training requests and relevant APIs ($\mathbb{X,Y}$), pretrained LLM weights ($\theta_0$),
number of stochastic generations ($m$), 
minimum draft reward ($r_{min}$), top reward percentile threshold ($P_{top}$), 
number of top drafts to keep per sample ($n_{draft}$)}
\Output{the final trained model ($\theta_T$)}

\Comment*[l]{for each alignment iteration}
\For{t in $1 \dots T$}
{
    \Comment*[l]{generate queries for the train dataset, sample m times}
    $\mathbb{\widehat{Z}}_{1..m} \leftarrow \text{generate\_queries}(\mathbb{X}, \theta_{t-1}, m)$ \;
    \Comment*[l]{use queries in retrieval and compute rewards}
    $\mathbb{R}_t \leftarrow \text{compute\_rewards}(\mathbb{X}, \widehat{\mathbb{Z}}_{1..m}, \mathbb{Y})$\; 
    \Comment*[l]{filter on min reward and top-percentile}
    $\mathbb{X}_t, \mathbb{Z}_t \leftarrow \text{filter\_samples}(\mathbb{X}, \widehat{\mathbb{Z}}_{1..m}, \mathbb{R}_t, r_{min}, p_{top}, n_{draft})$ \;
    \Comment*[l]{supevised fine-tuning on filtered generations}
    $\mathbb{\theta}_t \leftarrow \text{supervised\_fine-tuning}(\mathbb{X}_t, \mathbb{Z}_t, \theta_{t-1})$ \;
}
\end{algorithm}

Algorithm~\ref{alg:alignment} shows an overview of this process. We start from a pre-trained LLM, then for T alignment iterations, use the model from the most recent iteration to generate a set of $m$ queries ($\mathbb{\widehat{Z}}_{1..m}$) for each training sample and relevant API pair ($\mathbb{X,Y}$). To generate such queries given the most recent iteration of the model $\theta_{t-1}$, we use stochastic generation to promote diversity among the generated drafts. Then, we simulate retrieval of items in the train set using the generated queries in $\mathbb{\widehat{Z}}_{1..m}$ and compute retrieval reward for all samples. A simple filter is applied on the reward values to only keep the top $n_{draft}$ generated query sets (drafts) with the highest rewards
, and subsequently remove any remaining draft that has a reward value less than $r_{min}$ or falls outside the $p_{top}$ percentile of the population. Finally, we finetune the model on the filtered samples i.e. request and generated queries using similar settings as in Supervised fine-tuning. This process is repeated T times to iteratively improve the model’s capability to generate better queries.

Regarding the reward metric, we experimented with MMRR, MAP, and average recall. While the choice of reward is use-case specific, we observed the best results for MMRR as the reward metric (see Appendix~\ref{sec:appendix:reward_metric}).

\section{Experiments}
\subsection{Dataset}
For our experiments, we used the dataset published by \citet{qin2023toolllm} 
which has requests and relevant APIs covering complex and multi-tool scenarios. We conducted a simple preprocessing step to reduce low-quality API documents and samples. Specifically, we remove API documents that have descriptions that are shorter than $5$ words or longer than $50$ words as well as samples with no relevant API assignment or more than $3$ APIs assigned. This preprocess step results in a smaller set of about $1,831$ APIs. 

Subsequently, we split the APIs into $1,458$ in-domain and $373$ out-of-domain sets randomized based on tool names. The in-domain set is further divided into $15,987$ training and $1,776$ in-domain test requests. The out-of-domain test set consists of $4,451$ examples. During the split process, to ensure a complete split and no contamination between in-domain and out-of-domain sets, we removed any sample that had relevant APIs overlapping the other set. Throughout this paper, we use the in-domain training set for experiments that require any form of training/supervision. The test datasets are only used for evaluation.

\subsection{Retriever Setup}
We focus our experiments on a retriever which builds an index on API descriptions. This retriever uses a set of queries during retrieval to efficiently find relevant APIs.
We use \textit{all-mpnet-base-v2}\footnote{\url{https://huggingface.co/sentence-transformers/all-mpnet-base-v2}}~\cite{reimers2019sentence} as the embedding model. The retrieval is done via a simple flat index and nearest neighbor search with cosine distance metric. To retrieve a ranked list based on a set of generated queries, we use an interleaving scheme. 
The interleaving method independently retrieves items based on each generated query sorted by the similarity metric. Then we iterate over the lists and take one item from each while skipping duplicates to compose the final retrieval result.







In our early experiments, we found that appending the original request to the set of generated queries generally improves the retrieval metrics. Therefore, for any experiment that involves query generation, we use this by default. For ablation study on the impact this method, please refer to Appendix~\ref{sec:appendix_ablation}.

\subsection{Query Generation Setup}
As introduced in Section~\ref{sec:LLM-Based Query Generation}, we experiment with four main cases: (a) the baseline setup of using user request as is for the retrieval referred to as Utterance, (b) leveraging an out-of-box LLM for query generation denoted by Zero-Shot, (c) fine-tuning the model for query generation on the training split requests/APIs (SFT), and (d) leveraging the alignment learning technique that iteratively improves the query generation capability without directly fine-tuning on API documents (Alignment). 

For each case, we conduct a basic temperature calibration by measuring the Recall@5 performance while varying the temperature in the range of $0$ to $1.7$ with increments of size $0.2$. More details on specific hyper-parameter settings is presented in Appendix~\ref{sec:appendix_hyperparameters}.

\subsection{Results}
Table~\ref{tab:main_results} presents a comparison of the results. For the in-domain test set, SFT results in the best retrieval metrics. However, for the out-of-domain scenario, the alignment method consistently shows the most promising results. This result suggests that for applications that require supporting out-of-domain APIs, the alignment approach is more promising. Note that for many practical applications due to the cost of LLM training, it is not feasible to retrain the model when dealing with a growing number of new APIs.

\begin{table}[t]
\centering
\resizebox{\linewidth}{!}{
\renewcommand{\arraystretch}{1.3}
\begin{tabular}{l|cccc}
\toprule
\textbf{Metric} & \textbf{No Gen.} & \multicolumn{3}{c}{\textbf{LLM-Gen.}} \\
& \textbf{Utterance} & \textbf{Zero-Shot} & \textbf{SFT} & \textbf{Alignment} \\
\hline
& \multicolumn{4}{c}{\textbf{In-Domain Evaluation}} \\
\textbf{MMRR} & 0.4841 & 0.4145	& \textbf{0.7370}	& 0.6925 \\
\textbf{MAP} & 0.5675 & 0.5111	& \textbf{0.7508}	& 0.7225 \\
\textbf{Recall@3} & 55.67\% & 61.28\% & \textbf{80.95\%} & 76.36\% \\
\textbf{Recall@5} & 63.82\% & 57.86\% & \textbf{87.29\%} & 85.34\% \\
\textbf{Recall@11} & 74.36\% & 70.47\% & \textbf{91.60\%} & 91.00\% \\
\hline
& \multicolumn{4}{c}{\textbf{Out-Of-Domain Evaluation}} \\
\textbf{MMRR} & 0.6290 & 0.5440 & 0.6130 & \textbf{0.6487} \\
\textbf{MAP} & 0.7031 & 0.6432 & 0.6893 & \textbf{0.7151} \\
\textbf{Recall@3} & 69.68\% & 52.78\% & 68.56\% & \textbf{71.04\%} \\
\textbf{Recall@5} & 75.26\% & 71.76\% & 76.18\% & \textbf{78.53\%} \\
\textbf{Recall@11} & 82.79\% & 80.86\% & 82.83\% & \textbf{85.51\%} \\
\hline
\end{tabular}
}
\caption{Comparison of retrieval performance for the in-domain and out-of-domain evaluation sets using the user utterance as the retrieval query as well as LLM-based query generation methods including zero-shot prompting, SFT, and alignment learning.}
\label{tab:main_results}
\end{table}

\begin{figure}[t!]
    \centering
        \includegraphics[width=1.0\linewidth]{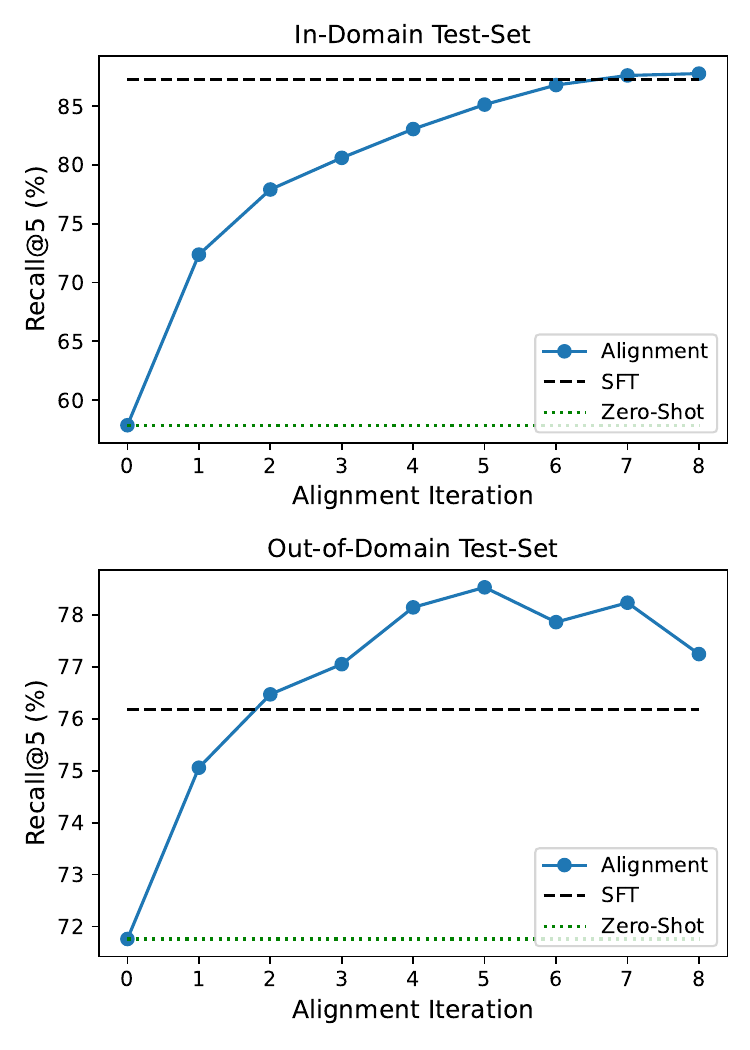}
        \caption{Comparison of Recall@5 performance for the zero-shot, SFT, and alignment iterations reported for the in-domain and out-of-domain evaluation sets.}
        \label{fig:alignment_iters}
    \centering
        \includegraphics[width=1.0\linewidth]{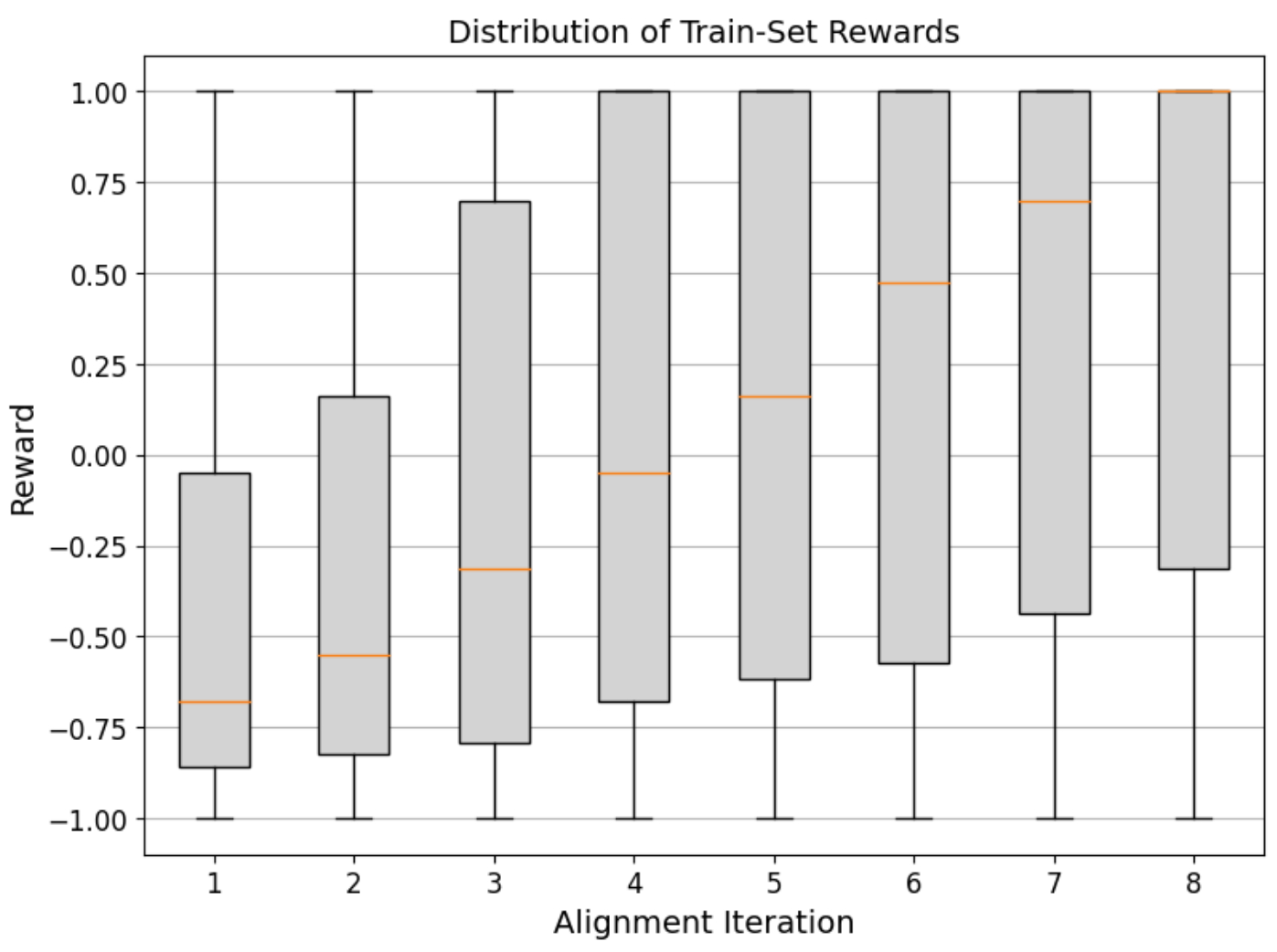}
        \caption{Bar plots showing the distribution of rewards for the train set examples during the iterative alignment process. With more iterations the reward distribution shifts significantly toward higher values.}
        \label{fig:barplot_train_reward}
\end{figure}

Figure~\ref{fig:alignment_iters} shows how Recall@5 evolves over the alignment iterations. In this case, the best out-of-domain performance is reached after $5$ iterations, while the in-domain performance is consistently improving. We found that with increasing the number of alignment iterations, the performance of this method surpasses SFT, however, usually at that point the out-of-domain performance starts to decline, potentially due to overfitting to the limited train set. While in the experiment results shared in Table~\ref{tab:main_results}, we do not evaluate models at such operating point 
and aim for the best out-of-domain performance, depending on the application, it could be a better balance to train for more iterations and enjoy a better in-domain performance at a marginal cost to the out-of-domain performance.

To dive deeper into the progression of rewards during the alignment process 
, we used bar plots in Figure~\ref{fig:barplot_train_reward} to show the distribution of at each iteration. From this figure, we can see the distribution of rewards measured on the train set monotonically increases with the alignment iterations. This figure indicates overfitting on the in-domain data after the $7$th iteration which is consistent with the Recall@5 trends presented in Figure~\ref{fig:alignment_iters}.

\section{Conclusion}
In this study, we investigated improving the tool retrieval performance for complex and contextual cases. We showed that leveraging LLM-generated queries provides an effective method to introduce contextual and common-sense understanding to the retrieval process. We experimented with different approaches such as zero-shot prompting, supervised fine-tuning, and alignment learning. Based on the experimental results, we found that alignment learning guides the LLM to generate queries that result in the best end-to-end retrieval performance, especially for the challenging out-of-domain settings where tools are not seen during the training process.

\bibliography{refs}

\clearpage

\appendix
\section{Hyperparameter Settings}
\label{sec:appendix_hyperparameters}
\subsection{Generation}

For each model, we conduct a basic temperature calibration by measuring the Recall@5 performance while varying the temperature in the range of $0$ to $1.7$ with increments of size $0.2$. We found that the best temperature for the evaluation of the Zero-Shot, SFT, and alignment methods is $1.3$, $0.6$, and $0.1$, respectively. For all cases, we consider top 90\% of the token distribution, and consider 10 highest probability tokens at each token generation step.

\subsection{Training}
For all experiments that require training, we use a batch size of 32, a constant learning rate of $2\times10^{-5}$, and set the weight decay to $0.01$. We use gradient clipping to clip values outside the range of $[-1,1]$. The loss is only computed for the generated tokens to prevent forcing the distribution of input/task tokens. For the SFT training or each iteration of alignment, we only train for one epoch as we found this to significantly reduce overfitting issues.

\subsection{Alignment}
\label{sec:appendix_Alignment}
For the alignment learning experiments, to generate training samples, we use a typical temperature of $1.0$ and generate $24$ drafts for each sample. Regarding the filter setup, we explored different reward metrics and values for $P_{top}$,$r_{min}$, and $r_{draft}$, but found best results for using MMRR, $P_{top}=100$,$r_{min}=0.05$, $r_{draft}=1$, and $T=5$. Note that due to the computational cost of these experiments, we were not able to cover a complete grid search space to find the optimal settings, and instead limited search space by finding a reasonable working setting and changing variables one at a time. 

\section{Ablation Study}
\label{sec:appendix_ablation}
\subsection{Impact of Adding Utterance to the Query Set}
Table~\ref{tab:ablation_utt_add} shows ablation results for the change in performance when the original utterance is not added to the query set. As it can be seen, including the original utterance in the queries used for retrieval consistently helps the zero-shot prompting method, especially for the case of out-of-domain evaluation. However, For the SFT and alignment learning methods, we do observe some regressions for the in-domain evaluation set. This is likely due to the capability of these models to generate queries that are of enough quality so is beneficial to solely rely on them. Nevertheless, since the gains for the case of out-of-domain evaluation are more significant, we decided to consistently append the original utterance to the query set for the main results presented in this paper. Based on the presented results, such decision may need to be revisited for use-cases that are only interested in the in-domain performance.

\begin{table}[t]
\centering
\resizebox{\linewidth}{!}{
\renewcommand{\arraystretch}{1.3}
\begin{tabular}{l|ccc}
\toprule
\textbf{Metric} & \multicolumn{3}{c}{\textbf{Delta wrt. Adding Utterance to the Query Set}} \\
& \textbf{Zero-Shot(-utt)} & \textbf{SFT(-utt)} & \textbf{Alignment(-utt)}  \\
\hline
& \multicolumn{3}{c}{\textbf{In-Domain Evaluation}} \\
\textbf{MMRR} & -0.0868  & 0.0286 & 0.0114  \\
\textbf{MAP} & -0.1330  & 0.0452  & 0.0201  \\
\textbf{Recall@3} & -13.42\%  & -1.41\%  & 0.34\%  \\
\textbf{Recall@5} & -8.67\%  & -4.54\%  & 1.91\%  \\
\textbf{Recall@11} & -8.58\%  & -5.52\%  & -1.87\%  \\
\hline
& \multicolumn{3}{c}{\textbf{Out-Of-Domain Evaluation}} \\
\textbf{MMRR} & -0.0863 & -0.1766  & -0.0669  \\
\textbf{MAP}	& -0.1292  & -0.2095 & -0.0719  \\
\textbf{Recall@3} & -14.50\%  & -19.77\%  & -6.42\%  \\
\textbf{Recall@5} & -7.84\% & -21.28\% & -6.36\%  \\
\textbf{Recall@11} & -6.09\%  & -19.38\% 	& -4.65\% \\
\hline
\end{tabular}
}
\caption{Ablation study on the impact of adding the original utterance to the query set. Delta values reporting compared to the default case of adding the utterance.}
\label{tab:ablation_utt_add}
\end{table}

\subsection{Impact of Changing Sample Filtering Configurations}
The rejection sampling method used for alignment learning can be particularly sensitive to filter settings as it needs to remove low-reward responses while ensuring diversity in the produced training samples. As explained in Section~\ref{sec:appendix_Alignment}, we conducted experiments for finding the right hyperparameter settings for the alignment learning method.
See Table~\ref{tab:change_filter_params} on the impact of changing rejection sampling filter configurations. While additional investigation is required to find optimal settings for new model architectures and datasets, we found that generating as many as $24$ response drafts, keeping the one with highest reward, and filtering out any sample that has very low reward generally results in stable convergence and outperforming alternative methods.

\begin{table}[t]
\centering
\resizebox{\linewidth}{!}{
\renewcommand{\arraystretch}{1.3}
\begin{tabular}{lccc|cc}
\toprule
\multicolumn{4}{l}{\textbf{Experiment}} & \multicolumn{2}{c}{\textbf{Recall@5 Delta wrt. Baseline}} \\
& $p_{top}$	& $r_{min}$	& $n_{draft}$ & In-domain & Out-Of-Domain \\
\hline
Baseline & 100 & 0.05 & 1 & 0\% & 0\% \\
Reduced $p_{top}$ &	75 &0.05 & 1 & -0.28\% & -0.29\% \\
Increased $r_{min}$	& 100	& 0.3 & 1 & -0.50\% & -0.01\% \\
Decreased $r_{min}$	& 100 & 0 & 1 & -0.06\% & -0.05\% \\
Increased $n_{draft}$ & 100	& 0.05 & 2 & -0.29\% & -1.43\% \\
\hline
\end{tabular}
}
\caption{Impact of changing rejection sampling filter hyper-parameters. Delta values are reported compared to the baseline of: $p_{top}=100\%, r_{min}=0.05, n_{draft}=1$.}
\label{tab:change_filter_params}
\end{table}

\subsection{Impact of Changing the Reward Metric}
\label{sec:appendix:reward_metric}
Table~\ref{tab:change_reward_metric} presents a comparison of Recall@5 results for using different retrieval reward metrics i.e. MMRR, MAP, and average recall. While choosing a reward metric is use-case specific, we decided to use MMRR as it provides a more intuitive measure of retrieval recall quality compared to the MAP. Compared to leveraging recall average as the reward metric, MMRR provides a more smooth target that encourages better retrieval for all positions rather than focusing on a fixed cut-off.

\begin{table}[t]
\centering
\resizebox{\linewidth}{!}{
\renewcommand{\arraystretch}{1.3}
\begin{tabular}{l|cc}
\toprule
\textbf{Reward Metric} & \multicolumn{2}{c}{\textbf{Recall@5 Delta wrt. MMRR Reward	}} \\
  & In-domain & Out-Of-Domain \\
\hline
MMRR & 0\% & 0\% \\	
MAP & -3.45\% & -0.82\% \\	
Avg(Recall@5,Recall@11) & -0.87\% & -1.10\% \\
\hline
\end{tabular}
}
\caption{Impact of changing the alignment learning reward metric. Delta values reported compared to the default case of MMRR as the reward metric.}
\label{tab:change_reward_metric}
\end{table}

\section{Generating Intent vs. Description as Query}
\label{sec:appendix_itent}
Intent prediction is a classical approach to user understanding and skill selection in dialogue systems. Compared to tool descriptions, intents represent a different level of abstraction and potentially reduce some of the hallucination models such as made-up tool names. To evaluate the impact of generating intents rather than tool descriptions, we used a new prompt to instruct the model to generate a list of intent descriptions as queries. See Figure~\ref{fig:prompt_intent_gen}. Except for this change, we used the exact same process to train and evaluate the alignment learning method.

\begin{figure}[t]
\begin{mdframed}[backgroundcolor=prompt_background,linewidth=0.0pt]
\texttt{\small
Given a request by user (Human), generate the description of the user's intentions (intents).\\
Try to decompose the request to a set of intents.\\
Do NOT respond to the Human and just describe the intents.\\
Use new line to separate multiple descriptions. Each description should be less than 20 words.\\
Return at most 5 descriptions (lines).\\
Do not provide any additional explanation or examples, return just a set of intents.\\
\\
Human: <user request>
\\
Answer:
}
\end{mdframed}
\caption{Prompt format used for the intent generation experiments.}
\label{fig:prompt_intent_gen}
\end{figure}

Table~\ref{tab:intent_gen} shows a comparison of results for the alignment learning method when changing the prompt format and generating intent descriptions as the retrieval query. Overall, the tool description generation appears to perform marginally better. However, we believe that intent description is an interesting direction for future work and the two methods can potentially complement each other.

\begin{table}[t]
\centering
\resizebox{\linewidth}{!}{
\renewcommand{\arraystretch}{1.3}
\begin{tabular}{l|cc}
\toprule
\textbf{Reward} & \multicolumn{2}{c}{\textbf{Recall@5 Delta wrt. Description Generation}} \\
 & Description Generation & Intent Generation \\
\hline
  & \multicolumn{2}{c}{{In-Domain Evaluation}} \\
MMRR & 0 & -0.0071 \\
Recall@5 & 0\% & -0.35\% \\
Recall@11 & 0\% & -0.62\% \\
\hline
 & \multicolumn{2}{c}{{Out-Of-Domain Evaluation}} \\
MMRR & 0 & -0.0040 \\
Recall@5 & 0\% & -0.29\% \\
Recall@11 & 0\% & -0.50\% \\
\hline
\end{tabular}
}
\caption{Impact of generating user intents as retrieval query rather than tool descriptions for the alignment learning method. Delta value are reported compared to the tool description generation approach.}
\label{tab:intent_gen}
\end{table}

\clearpage
\section{Qualitative Results}
Table \ref{tab:qualitative} presents the outputs of our alignment learning query generation model for a set of entirely unseen authored examples. To author these examples, we did not consider the pool of available APIs in the training set or even the distribution/structure of data seen during the alignment process. For all of the tested cases in this table, the model is able to reduce ambiguity and complexity by breaking-down the original request to a set of simpler more direct tool descriptions. 

In the second example, the generated queries show common-sense understanding capabilities by focusing on suggestions for outdoor activities. In the third example, the generated query list is a decomposition of the initial request which usually facilitates retrieval for semantic retrievers. One behavior that is noticeable in the examples here is that the generated outputs contain hallucinated API names such as "PartyPlanningAPI" which points to non-existent APIs. While we did not observe a noticeable performance regression for such cases and the generated API names are often contain relevant keywords, we believe there is room for further improvement to address this hallucination mode.

\begin{table*}[b]
\centering
\resizebox{\linewidth}{!}{
\renewcommand{\arraystretch}{1.3}
\begin{tabular}{l|l}
\toprule
\textbf{Request} & \textbf{Generated Queries} \\
\hline
Give me a good advice for making coffee 
  & - Coffee Tips: Offers a list of tips for making coffee \\
  &  - Coffee Machine Reviews: Offers reviews and ratings on different coffee machines \\
  &  - Coffee Beans: Provides information on different types of coffee beans \\
  &  - Coffee Brewing: Provides tips and techniques for brewing coffee \\
\hline
I'm bored and tired of staying home. Literally 
 & - Get a list of nearby parks and recreational areas\\
watched tv all day. Give me some ideas what to do & - Search for nearby places to visit based on location and interests \\
& - Get a list of nearby restaurants and cafes, including reviews and ratings \\
& - Get a list of popular tourist attractions and activities, including theme parks, \\
& museums, and other tourist attractions \\
\hline
I'm planning for a small party and need to invite guests, 
& - PartyPlanningAPI: Create a party planning checklist for cleaning, decorating, and ordering food \\
clean the house, and order food. Can you help me? & - HouseCleaningAPI: Get a list of cleaning tasks for the house \\
& - FoodOrderAPI: Place orders for food delivery or pickup \\
& - GuestListAPI: Create a list of guests and their contact information \\
\hline
I want to get more healthy, where should I start?
& - Nutrition API: Offers a range of nutritional information and tools to help you make healthy choices \\
& - Healthy Eating API: Provides a list of healthy foods and drinks to help you make healthy choices \\
& - Medication API: Offers information on various medications and their effects on your health \\
& - Healthy Living API: Provides a range of healthy living tips and resources \\
& to help you improve your overall health \\
\hline
\end{tabular}
}
\caption{Examples of authored out-of-domain test requests and generated queries by the alignment learning model. Generated queries generally decompose the request, eliminate unnecessary explanations, and show common-sense contextual understanding.}
\label{tab:qualitative}
\end{table*}

\end{document}